\documentclass{article}

\usepackage[preprint]{neurips_2026}

\usepackage[utf8]{inputenc} 
\usepackage[T1]{fontenc}    
\usepackage{hyperref}       
\hypersetup{hypertexnames=false}
\usepackage{url}            
\usepackage{graphicx}       
\usepackage{booktabs}       
\usepackage{multirow}       
\usepackage{amsfonts}       
\usepackage{amsmath}        
\usepackage{algorithm}      
\usepackage{algpseudocode}  
\usepackage{nicefrac}       
\usepackage{microtype}      
\usepackage{xcolor}         
\usepackage{caption}        
\newcommand{\method}{\textit{TaintedPixels}}

\title{Can Tainted Pixels Expose Deepfake Videos?}

\author{%
  Juan Hu \quad
  Shaojing Fan \quad
  Sanjay Saha \quad
  Marc Herrera \quad
  Terence Sim \\[2mm]
  National University of Singapore
}

\begin{document}

\maketitle

\begin{abstract}
Publicly-accessible face-manipulation tools have made deepfake creation accessible 
to non-expert users. Against these, existing defenses are mostly post-hoc, detecting only after 
forgery has occurred, and operating on still images rather than videos. Research is lacking in (i) the proactive protection of published facial videos against 
black-box manipulation tools, and in (ii) understanding its perceptual effect on human 
viewers. We introduce \method{}, a proactive 
video-protection method built around an asymmetric visibility 
trade-off: the embedded watermark should remain inconspicuous in the 
published video but become obvious once a downstream tool 
manipulates the video. \method{} injects structured periodic perturbations 
into the blue channel of facial regions and refines them under 
stripe-visibility, color-cast, and video-level LPIPS budgets, with 
lightweight motion-adaptive deployment. We believe \method{} is
the first proactive defense designed specifically against  
black-box manipulation tools rather than image-level pipelines or specific surrogate generators. Across three publicly available off-the-shelf video manipulation tools and two 
off-the-shelf detectors, \method{} attains the highest forgery fake 
rate while keeping perturbations small (LPIPS = 0.0042). Our non-expert human study, conducted on a diverse set of 400 video stimuli spanning different lighting conditions, backgrounds, and skin tones, shows that protected source videos draw a 3.26\% suspicion rate, while forgeries from protected sources are 
identified as fake much more often than forgeries from 
unprotected sources (90.72\% vs.\ 56.71\%). This validates the effectiveness of \method{}. 
\end{abstract}

\section{Introduction}

Recent advances in face swapping and audio-driven facial animation 
have made deepfake video synthesis both highly realistic and broadly 
accessible (\cite{syed2026exploring,baliah2025realistic,hong2025audio,
ki2025float,wang2025dynamicface}). These same capabilities, while useful for entertainment and virtual avatars, significantly lower the barrier to identity misuse, enabling convincing impersonation and the rapid spread of misinformation at scale 
(\cite{alanazi2025unmasking,diel2025harm,diel2024human}). The threat 
is no longer confined to expert-operated research systems: public 
implementations such as Roop, FaceFusion, and MuseTalk allow 
non-expert users to perform face swapping or lip-synchronized 
animation with minimal technical effort 
(\cite{roop2023github,facefusion2026github,musetalk2026github}). This poses a severe threat to public figures, whose abundant online appearances provide ready training material for high-fidelity forgeries that can be weaponized for political disinformation, financial fraud, and reputational harm (\cite{chesney2019deepfakes}).

Defenses against this misuse fall into two broad families. 
\emph{Post-hoc} approaches detect manipulated content after the fact 
(\cite{yan2024lsda,cui2025forensicsadapter,cai2025deepshield, Saha_2023_ICCV}); they have demonstrated promising results, but they intervene only after a forgery has been disseminated. \emph{Proactive} approaches instead 
modify benign media before the forgery, embedding either verifiable 
watermarks for provenance (\cite{zhao2023identitywatermarking,wang2024lampmark,wu2024advmark,xia2025towards}) or perturbations that 
disrupt the forgery model itself (\cite{qu2024dfrap,
jeong2025faceshield,wang2025nullswap,wang2025faceswapguard}). Compared to post-hoc methods, proactive approaches are becoming increasingly critical: once a forgery spreads online, takedown is often slow and incomplete, and cannot reverse the harm already inflicted; by contrast, proactive defenses act at the source, thwarting forgeries before they can be generated and widely circulated (\cite{nguyen2025survey}).

\textbf{Our solution.}
Any effective protection method
must satisfy an \emph{asymmetric visibility trade-off} with three key requirements: (i) the embedded perturbation should be imperceptible in the original video, preserving its visual quality and intended function; (ii) once processed by a manipulation pipeline, the signal should be reliably exposed to machine detectors; and (iii) the resulting forgery should appear perceptibly unconvincing or suspicious to human viewers. These objectives pull in different directions. We take a holistic view whereas previous works typically focus only on one direction.

\begin{figure*}[t]
\centering
\includegraphics[width=\textwidth]{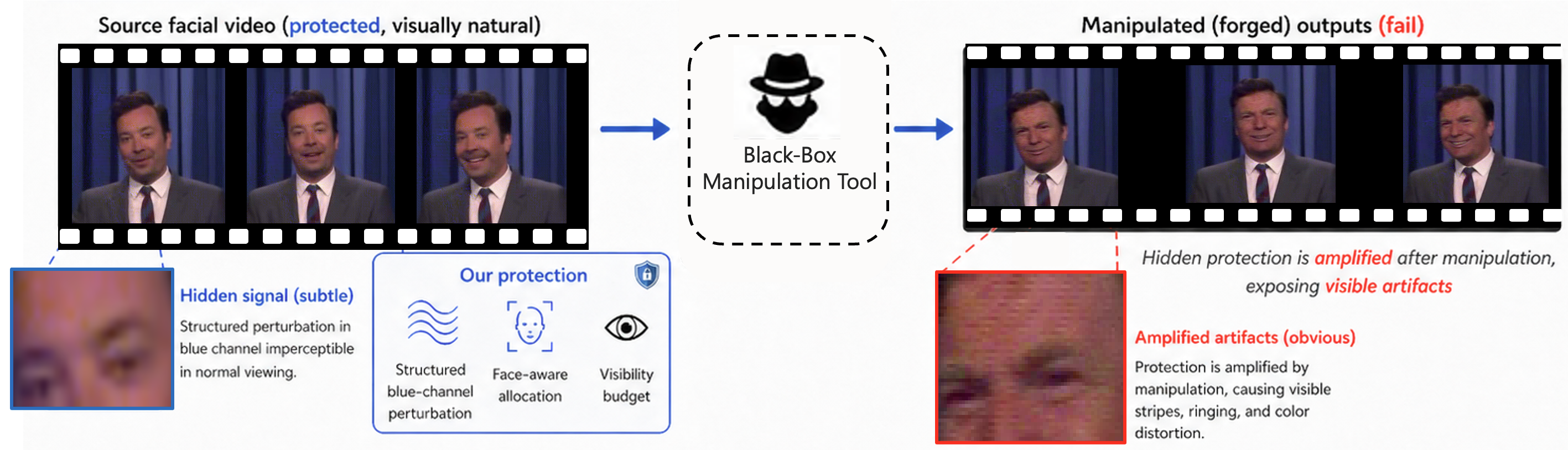}
\caption{Inspired by anti-copy patterns in passports, \method{} embeds an invisible signal into a source video, protecting it. Visible artifacts appear if the video is manipulated.}
\label{fig:fig1}
\end{figure*}

\paragraph{\method{}.}
We instantiate this principle in \method{}, a visibility-budgeted 
proactive video-protection framework (Figure~\ref{fig:fig1}). 
\method{} places structured periodic perturbations in the blue 
channel of facial regions, where human chromatic sensitivity is 
lowest (\cite{mullen1985contrast,cole1993detection,curcio1991distribution}), and refines them under three explicit budgets: a 
stripe-visibility budget, a color-cast budget weighted toward skin 
and bright regions, and a video-level LPIPS budget that bounds 
perceptual deviation across frames. A lightweight motion-adaptive scaling step adapts the signal to fast-moving regions without 
redoing optimization. 

Importantly, our method requires no access to, or make any assumptions about, the downstream manipulation tool; the protection is generated solely from the source video under perceptual constraints. This design enables a broader, public-facing defense mechanism that operates at the data level, aiming not only to support detectors but also to reduce the credibility of deepfakes at scale before they can be convincingly synthesized and circulated. The perturbations from \method{} are designed to remain imperceptible to human viewers, ensuring that the protected video preserves its visual quality and remains suitable for general use and distribution.

\paragraph{Threat model and scope.}
We envisage this scenario: a user publishes a protected video and an attacker applies an off-the-shelf manipulation tool on it, unaware that it is protected. We evaluate three representative tools 
spanning two manipulation paradigms: Roop (\cite{roop2023github}) and 
FaceFusion (\cite{facefusion2026github}) for face swapping, and 
MuseTalk (\cite{musetalk2026github}) for audio-driven animation. We 
do not claim universal robustness against future, closed-source, or 
adaptive forgery models. We ask the more circumscribed question of 
whether proactive protection can render forgeries from current 
public tools visibly unreliable while keeping the published video 
natural.

\textbf{Our contributions:} (i) We propose \method, a novel deepfake defense which protects videos \emph{at the source}, to facilitate detection of subsequent manipulation; (ii) we report the first-ever human perceptual study showing that protected videos, when forged, are easier to detect by the naked eye.

\section{Related Work}
\label{sec:related_work}

Due to space constraints, we will not review work related to deepfake synthesis or detection. Interested readers may refer to \cite{liu2025review}. Instead, we will focus on research that is closer to ours.



\paragraph{Disruption-based proactive defense.}
Within proactive defenses, disruption-based methods perturb benign media so that
downstream manipulation models fail to synthesize realistic outputs. Early work
studies adversarial perturbations against facial manipulation systems
(\cite{ruiz2020disrupting}), while subsequent methods improve robustness,
transferability, efficiency, or perceptual quality under transformations,
compression, or specific model families
(\cite{aneja2022tafim,wang2022anti,wang2022deepfake,qu2024dfrap,
zhu2024facepoison,jeong2025faceshield,huang2022cmua}). Recent identity-cloaking methods further
protect facial identity under black-box face swapping by perturbing identity
features or weakening identity transfer (\cite{wang2025nullswap,
wang2025faceswapguard, 10943380}).

Our work follows the disruption-based line but differs from prior work in both
setting and objective. Instead of optimizing image-level perturbations against a
specific surrogate generator or focusing primarily on identity suppression, we
insert video-level protection against off-the-shelf black-box manipulation tools.
This shifts the objective from suppressing identity transfer to exposing
manipulation-induced artifacts after the downstream pipeline. Accordingly, our
perturbation design emphasizes source-side perceptual budgeting and video-level
robustness, rather than generic norm-bounded noise. We instantiate this idea with
structured chromatic perturbations motivated by human color sensitivity
(\cite{mullen1985contrast,curcio1990human}) and constrain perceptual deviation
using learned perceptual similarity (\cite{zhang2018perceptual}).

\paragraph{Watermarking and provenance-based defenses.}
Another proactive approach embeds verifiable signals into facial media for
authentication, attribution, provenance tracking, or proactive detection.
Representative methods include recoverable provenance tags, identity
watermarks, landmark perceptual watermarks, detector-aware adversarial
watermarks, and contour-hybrid watermarks
(\cite{wang2021faketagger,zhao2023identitywatermarking,wang2024lampmark,
wu2024advmark,xia2025towards}). These methods are useful for verification and
source tracing, but they usually require a watermark extraction or detection
stage. \method{} instead treats the embedded signal as a disruption source
rather than a recoverable provenance message. No watermark detector is needed.



\paragraph{Human perception and social media dynamics.}
Human perception and online dissemination jointly motivate proactive defenses. People often struggle to distinguish manipulated content, especially when it appears realistic (\cite{diel2024human,groh2022deepfake,koebler2025deepfake}), so a forgery that looks plausible can still mislead viewers even if detected by machines. At the same time, manipulated media spread rapidly and widely, while corrections rarely reach the same audience (\cite{vosoughi2018spread,shao2018spreadinglowcredibility}), meaning takedowns often come too late (\cite{chesney2019deepfakes}). These findings motivate two design choices in \method: protection should be embedded \emph{at the source}, and its effect should appear as \emph{visible artifacts} in forged outputs, analogous to anti-copy patterns in secure documents that only emerge upon reproduction (\cite{garg2023analysis}).

\section{Method}
\label{sec:method}

\subsection{Overview}
\label{sec:method_overview}

We formulate proactive facial video protection as a visibility-budgeted disruption
problem. Given a clean facial video $\mathbf{V}=\{x_t\}_{t=1}^{T}$, where each
frame $x_t\in[0,1]^{H\times W\times 3}$, our objective is to produce a protected
video $\hat{\mathbf{V}}=\{\hat{x}_t\}_{t=1}^{T}$ that satisfies two asymmetric
requirements. Before manipulation, the protected video should remain visually
natural to human viewers. After manipulation, the same hidden protection should
be amplified by publicly available off-the-shelf tools and make the forged output visually
unreliable or easier to detect as fake. \emph{We emphasize that our method treats the manipulation tool as a black-box.} We do not use the tool's internals; we do not even require API access to the tool to implement our protection.

\begin{figure}[t]
  \centering
  \includegraphics[width=\textwidth]{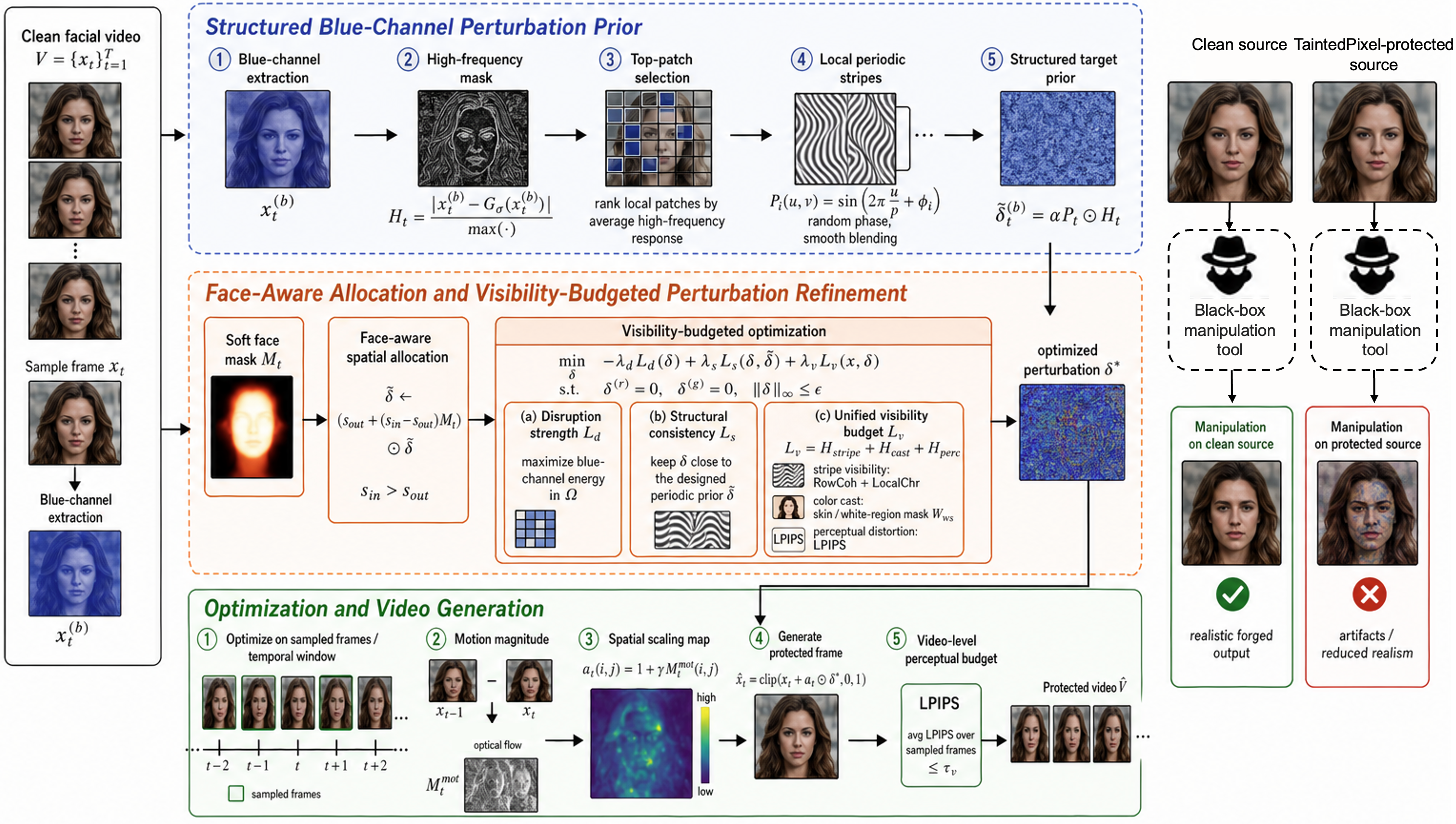}
  \caption{Overview of our proposed visibility-budgeted facial video protection framework.}
  \label{fig:method_overview}
\end{figure}
\vspace{-2mm}
As shown in Figure~\ref{fig:method_overview}, our framework contains three components. First, we construct a structured
blue-channel perturbation prior that concentrates energy in locally textured
regions. Second, we use a face-aware spatial prior to allocate more perturbation
budget to regions that manipulation tools are likely to process. Finally, in the optimization and video generation stage, we optimize the perturbation under visibility budgets and generate protected videos using lightweight motion-adaptive scaling with a video-level LPIPS constraint.


\subsection{Structured Blue-Channel Perturbation Prior}
\label{sec:blue_prior}

We restrict the perturbation to the blue channel and use $\mathrm{clip}(z,0,1)$ to denote element-wise clipping of pixel values
to the valid image range $[0,1]$: 
$  \delta_t = [0,0,\delta^{(b)}_t],
    \quad
    \hat{x}_t = \mathrm{clip}(x_t+\delta_t,0,1)$. 
    
This design is motivated by the asymmetric sensitivity of human perception and
downstream image processing. Humans are relatively less sensitive to small
blue-channel variations (\cite{mullen1985contrast,cole1993detection,curcio1991distribution}), yet face manipulation pipelines still process these
variations through normalization, alignment, feature extraction, and rendering (\cite{rossler2019faceforensics++, li2020faceshifter, tolosana2020deepfakes, mirsky2021creation,walczyna2023quick}). Thus, a small structured chromatic signal can remain unobtrusive in the published source video but become amplified by manipulation operations.

 We construct a structured target perturbation $\tilde{\delta}_t$ that combines local
periodic patterns with a high-frequency content mask, so that the protection signal is preferentially placed in locally textured regions where small chromatic variations are less perceptible while still retaining disruption potential after manipulation (\cite{mullen1985contrast,ferwerda1997model}): $ 
\quad    \tilde{\delta}^{(b)}_t = \alpha\, P_t \odot H_t .$ 

Here, $H_t\in[0,1]^{H\times W}$ is a high-frequency mask, $P_t$ is a local
periodic pattern, and $\odot$ denotes element-wise
(pixel-wise) multiplication. The mask $H_t$ is computed from the blue channel by subtracting
a Gaussian-smoothed version:
\begin{equation}
    H_t =
    \frac{\left|x^{(b)}_t - G_{\sigma}(x^{(b)}_t)\right|}
    {\max\left|x^{(b)}_t - G_{\sigma}(x^{(b)}_t)\right|+\xi};
    \qquad 
     P_i(u,v)=\sin\left(\frac{2\pi}{p}u+\phi_i\right).
     \label{eq:H_P}
\end{equation}
where $G_{\sigma}$ denotes Gaussian smoothing, and $\xi$ is a small numerical constant for stable division. We then rank local patches
according to their average high-frequency response and inject periodic signals
only into the top fraction of patches. This concentrates perturbations in
visually complex regions, where small changes are less noticeable.

For a selected local patch, the periodic pattern, $ P_i(u,v)$, is defined as shown in Eq.~\eqref{eq:H_P},
where $p$ is the stripe period and $\phi_i$ is a random phase. Overlapping patches
are blended by a smooth window to avoid block boundaries. This design enforces spatial coherence in the perturbation. Compared to independent random noise, such coherent patterns are more robust to rendering transformations and are thus more likely to manifest as structured artifacts after manipulation.

\subsection{Face-Aware Allocation and Visibility-Budgeted Perturbation Refinement}
\label{sec:face_aware}

\subsubsection{Face-Aware Allocation}
Public face manipulation tools do not process all pixels equally. Face swapping
and face animation models usually crop, align, normalize, and re-render the
facial region, while background pixels are often ignored or weakly processed (\cite{qu2024dfrap,wang2025nullswap}).
Under a fixed visibility budget, perturbation energy placed in the background is
therefore less useful than the perturbation energy placed on the face.

To exploit this observation, we estimate a soft face mask
$M_t\in[0,1]^{H\times W}$ and use it to reweight the target perturbation:
\begin{equation}
    \tilde{\delta}^{(b)}_t
    \leftarrow
    \left(s_{\mathrm{out}}+(s_{\mathrm{in}}-s_{\mathrm{out}})M_t\right)
    \odot \tilde{\delta}^{(b)}_t .
\end{equation}
Here $s_{\mathrm{in}}$ and $s_{\mathrm{out}}$ control the perturbation scale
inside and outside the face region, respectively. We use a soft mask rather than
a hard one to avoid visible discontinuities along the facial boundary.
In practice, the mask can be obtained from landmarks or face detectors; when
landmarks are unavailable, a coarse face region is sufficient because the
subsequent visibility constraints refine the perturbation.

This face-aware allocation does not assume access to the forgery model. It only
encodes a weak semantic prior: facial regions are more likely to be transformed
by downstream tools. This prior is particularly useful under black-box conditions,
where gradients or internal features of the manipulation tool are unavailable.

\subsubsection{Visibility-Budgeted Perturbation Refinement}
The structured prior provides a useful initialization, but directly applying it
may either be too weak to affect manipulation or too visible in the source video.
We therefore refine the perturbation by solving a visibility-budgeted optimization
problem. The key principle is to maximize disruption potential while penalizing
only the parts of the perturbation that exceed content-adaptive perceptual
budgets.

For each optimized frame (or a sampled temporal window), we solve
\begin{equation}
\label{eq:main_objective}
\begin{aligned}
\min_{\delta}\quad
& - \lambda_{\mathrm{d}}\mathcal{L}_{\mathrm{d}}(\delta)
+ \lambda_{\mathrm{s}}\mathcal{L}_{\mathrm{s}}(\delta,\tilde{\delta})
+ \lambda_{\mathrm{v}}\mathcal{L}_{\mathrm{v}}(x,\delta), \\
\text{s.t.}\quad
& \delta^{(r)} = 0,\quad \delta^{(g)} = 0,\quad
\|\delta\|_{\infty} \le \epsilon .
\end{aligned}
\end{equation}
Here, $\delta$ denotes the optimized perturbation and $\tilde{\delta}$ denotes
the structured target perturbation constructed from the blue-channel prior.
The three terms correspond to three goals: $\mathcal{L}_{\mathrm{d}}$ encourages
sufficient disruption strength, $\mathcal{L}_{\mathrm{s}}$ preserves the designed
spatial structure, and $\mathcal{L}_{\mathrm{v}}$ enforces source-side visual
quality. The constraints $\delta^{(r)}=\delta^{(g)}=0$ restrict the perturbation
to the blue channel, while $\|\delta\|_{\infty}\le\epsilon$ enforces a strict
magnitude bound.

\paragraph{Disruption term.}
This encourages the perturbation, $\mathcal{L}_{\mathrm{d}}(\delta)$, to use the blue-channel budget:
\begin{equation}
    \mathcal{L}_{\mathrm{d}}(\delta)
    =
    \frac{1}{|\Omega|}
    \sum_{(i,j)\in\Omega}
    \left(\delta^{(b)}(i,j)\right)^2 ;
    \qquad
     \mathcal{L}_{\mathrm{s}}(\delta,\tilde{\delta})
    =
    \frac{1}{|\Omega|}
    \sum_{(i,j)\in\Omega}
    \left(
    \delta^{(b)}(i,j)-\tilde{\delta}^{(b)}(i,j)
    \right)^2 .
    \label{eq:LdLs}
\end{equation}
where $\Omega$ denotes the support region of the structured perturbation.
Unlike conventional adversarial attacks that optimize against a specific model,
our setting assumes no access to the downstream manipulation pipeline. Therefore,
$\mathcal{L}_{\mathrm{d}}$ does not target a particular model, but instead acts
as a model-agnostic proxy objective. Intuitively, increasing the energy of the
structured chromatic signal within $\Omega$ increases the likelihood that it will
survive preprocessing and be amplified during manipulation.

\paragraph{Structural consistency.}
Keep the optimized perturbation $\delta$ close to the designed blue-channel prior $\tilde{\delta}$  via $\mathcal{L}_{\mathrm{s}}$ in Eq.~\eqref{eq:LdLs}.

This term prevents the optimization from degenerating into arbitrary
high-magnitude pixel noise. Without this constraint, maximizing
$\mathcal{L}_{\mathrm{d}}$ alone may produce perturbations that are less stable
under resizing, normalization, and rendering, and may also become more visible in
the source video. By preserving the local periodic structure of
$\tilde{\delta}$, $\mathcal{L}_{\mathrm{s}}$ encourages the optimized
perturbation to remain spatially coherent while still allowing sufficient
flexibility for visibility refinement.

\paragraph{Unified visibility budget.}
The visibility term constrains the source-side perceptual impact of the
optimized perturbation:
\begin{equation}
\label{eq:visibility_term}
    \mathcal{L}_{\mathrm{v}}(x,\delta)
    =
    \mathcal{H}_{\mathrm{stripe}}(x,\delta)
    +
    \mathcal{H}_{\mathrm{cast}}(x,\delta)
    +
    \mathcal{H}_{\mathrm{perc}}(x,\delta).
\end{equation}
It groups together the dominant visual failure modes caused by blue-channel
structured perturbations: visible stripe patterns, unnatural color cast, and
global perceptual distortion. Specifically, let $C_b(x)$ denote the blue-difference chroma component and $\quad   \Delta C_b = C_b(x+\delta)-C_b(x).$ 

We compute a stripe visibility score
\begin{equation}
    \mathcal{V}(x,x+\delta)
    =
    \beta_1\,\mathrm{RowCoh}(\Delta C_b)
    +
    \beta_2\,\mathrm{LocalChr}(\Delta C_b),
\end{equation}
where $\mathrm{RowCoh}(\cdot)$ measures row-wise coherent chroma changes and
$\mathrm{LocalChr}(\cdot)$ measures local chroma deviations. The corresponding
normalized hinge loss, $ \mathcal{H}_{\mathrm{stripe}}$,  is:
\begin{equation}
    \mathcal{H}_{\mathrm{stripe}}
    =
    \left[
    \frac{\mathcal{V}(x,x+\delta)}
    {\mathcal{B}_{\mathrm{vis}}(x)+\xi}
    -1
    \right]_+ ; \qquad
    \mathcal{C}(x,x+\delta)
    =
    \frac{
    \sum_{i,j} W_{\mathrm{ws}}(i,j)\,|\Delta C_b(i,j)|
    }{
    \sum_{i,j} W_{\mathrm{ws}}(i,j)+\xi
    } ,
    \label{eq:H_C}
\end{equation}
where $\mathcal{B}_{\mathrm{vis}}(x)$ is a content-adaptive visibility budget. To suppress color cast, especially in skin and bright white regions, we define a
soft white/skin mask $W_{\mathrm{ws}}$ and measure the masked chroma shift $ \mathcal{C}(x,x+\delta)$, defined in Eq.~\eqref{eq:H_C}. The color-cast hinge, $\mathcal{H}_{\mathrm{cast}} $, is
\begin{equation}
    \mathcal{H}_{\mathrm{cast}}
    =
    \left[
    \frac{\mathcal{C}(x,x+\delta)}
    {\mathcal{B}_{\mathrm{cast}}(x)+\xi}
    -1
    \right]_+ ; \qquad
     \mathcal{H}_{\mathrm{perc}}
    =
    \left[
    \frac{
    \frac{1}{|\mathcal{T}|}
    \sum_{t\in\mathcal{T}}
    \mathrm{LPIPS}(x_t,\hat{x}_t)
    }
    {\tau_v+\xi}
    -1
    \right]_+ ,
    \label{eq:H_H}
\end{equation}
where $\mathcal{B}_{\mathrm{cast}}(x)$ is a content-adaptive color-cast budget.

Finally, we constrain perceptual distortion $\mathcal{H}_{\mathrm{perc}} $ by averaging LPIPS over
a temporal window $\mathcal{T}$, as per Eq.~\eqref{eq:H_H}, 
where $\tau_v$ is the video-level perceptual budget. This term constrains the
average perceptual deviation across frames rather than a single optimized frame. Overall,
$\mathcal{L}_{\mathrm{v}}$ (Eq.~\eqref{eq:visibility_term}) penalizes only the perturbation components that exceed
their visibility budgets, allowing the method to preserve visual quality while
still maximizing useful disruption strength.

\subsection{Optimization and Video Generation}
\label{sec:optimization_generation}

We optimize the perturbation using Adam with projected updates. After each Adam
step, we project the perturbation back to the feasible blue-channel set:
$     \delta \leftarrow
    \Pi_{\|\delta\|_\infty\le\epsilon}(\delta),
    \quad
    \delta^{(r)}=\delta^{(g)}=0 .$ 

For efficiency, we optimize the structured perturbation on a sampled subset of
frames and propagate it to the full video. This strategy is motivated by the
temporal redundancy of facial videos: adjacent frames usually share similar
facial appearance, illumination, and background structure, so the same optimized
protection signal can be reused across neighboring frames instead of being
independently optimized for every frame.

To account for local temporal changes, we apply motion-adaptive perturbation
scaling during video generation. Specifically, we estimate a normalized motion
magnitude map $M_t^{\mathrm{mot}}\in[0,1]^{H\times W}$ between consecutive clean
frames using optical-flow magnitude. We then define
a spatial scaling map:
$ a_t(i,j)=1+\gamma M_t^{\mathrm{mot}}(i,j)$, 
where $\gamma$ is a small coefficient controlling the maximum additional
perturbation strength. Given the optimized perturbation $\delta^\star$, the
protected frame is generated as
$  \hat{x}_t =    \mathrm{clip}\left(x_t+a_t\odot\delta^\star,0,1\right)$. 

This deployment strategy reuses the same structured perturbation in relatively
stable regions while assigning slightly stronger protection to fast-moving
regions. The intuition is that temporal changes can mask fine-grained spatial
distortions, allowing fast-moving regions to carry a small amount of additional
protection without noticeably degrading source-video quality. The video-level
LPIPS constraint further regularizes this process by ensuring that the average
perceptual deviation across sampled frames remains within a video-level budget.

The overall procedure is practical for public-tool black-box protection because
the protection is generated entirely from the published source video and
perceptual constraints. It does not require gradients, intermediate features,
outputs, or internal parameters from the downstream manipulation tools. The
resulting protected video is then published. When an attacker
applies a public manipulation tool on it, the hidden structured perturbation should be
amplified by the tool's preprocessing and rendering pipeline, leading to visible
artifacts or reduced realism in the forged output.

\section{Experiments}
\label{sec:experiments}

\subsection{Experimental Setup}
\label{sec:exp_setup}

\paragraph{Public-tool black-box setting.}

We evaluate three representative off-the-shelf tools. For face swapping, we use
Roop (\cite{roop2023github}) and FaceFusion (\cite{facefusion2026github}). For audio-driven face animation, we use MuseTalk (\cite{musetalk2026github}). These
tools represent a realistic misuse scenario where non-expert attackers directly
apply public manipulation pipelines.


\textbf{Datasets and protocol.}
We evaluate on real facial videos sampled from FaceForensics++ (\cite{rossler2019faceforensics++}) and Celeb-DF (\cite{li2020celeb}), two representative datasets widely used in deepfake detection. FaceForensics++ provides 1,000 real videos and Celeb-DF provides 890 real videos. Although both datasets contain real and fake videos, we use only their real videos for the video-protection experiments. For each video, we
generate both a clean input and a protected input. Each manipulation tool is then
applied to both inputs, producing paired forged videos. All methods are
evaluated under the same manipulation and detection pipeline.

\textbf{Parameter setting. }We use $\sigma=1.0$ for the high-frequency mask, $96\times96$ overlapping
patches with initial stride $32$, and initial stripe period $p=2$. The face mask is obtained
from facial landmarks with a detector fallback and smoothed with
$\sigma_M=6.0$. We set $s_{\mathrm{out}}=0.8$ and select
$s_{\mathrm{in}}$ from a small range centered at $12$. The perturbation is
optimized with Adam for at most $120$ steps with learning rate $0.009$, followed
by projection to the feasible blue-channel set after each update. We tune the
main strength and visibility parameters by coordinate descent on the first
$N_{\mathrm{tune}}=6$ source frames using only source-side visibility and
disruption statistics, without accessing manipulation tools, detector outputs,
or human-study videos. All experiments are conducted on a single NVIDIA H100 GPU with 80GB of memory. The fake
manipulation pipeline runs at 1.03 frames per second, while generating a
protected video with \method{} takes 150.37 seconds per video on average.

\textbf{Metrics. }For source-side quality, we report PSNR, SSIM,  and video-level LPIPS
between $\mathbf{V}$ and $\hat{\mathbf{V}}$. For forgery-side exposure, we report both model-based and human-study results.
For model-based evaluation, because our forged videos are generated by
off-the-shelf tools, including MuseTalk, Roop, and FaceFusion, they act as
black-box inputs to the detectors and require generalization beyond specific
forgery pipelines. Given the limited availability of public detectors with strong generalization
ability and batch-evaluation support, we use two representative and relatively generalizable
detectors, NPR (\cite{tan2024rethinking}) and
Deep-Fake-Detector-v2-Model (\cite{Deep-Fake-Detector-v2-Model}), and report fake
rate and fake confidence. For human evaluation, we report the fake rate from the
human-study responses.

\subsection{Protection Effectiveness}
\label{sec:main_results}

\textbf{Comparisons. }We evaluate \method{} against proactive SOTA methods: DFRAP (\cite{qu2024dfrap}), CMUA (\cite{huang2022cmua}), FacePoison (\cite{zhu2024facepoison}), and FaceShield (\cite{jeong2025faceshield}). Table~\ref{tab:quality_matched_main_comparison} reports the main protection effectiveness under public black-box manipulation tools. Forgeries generated from our protected videos are more easily detected as such, compared to those protected by said SOTA methods. This shows that the hidden protection is manifested after manipulation. Although our method does not achieve the best PSNR or SSIM, it obtains a more favorable video-LPIPS trade-off, indicating that our protected videos are perceptually natural. Taken together, these results show that \method{} preserves strong source-side perceptual quality while still causing forged outputs to fail after manipulation.

\textbf{Robustness. } Published videos may undergo common post-processing before being manipulated. We
therefore evaluate robustness to H.264 compression and re-encoding. We
apply each post-processing operation to the protected video before manipulation
and report the fake rate of the resulting forged outputs. Results show that re-encoding and H.264 compression reduce the fake-detection accuracy by 3.13\% and 4.27\%, respectively. Nevertheless, \method{} helps preserve the protection signal under H.264 compression and re-encoding.

\begin{figure*}[h]
\centering
\begin{minipage}[t]{\textwidth}
\centering
{\includegraphics[width=\textwidth]{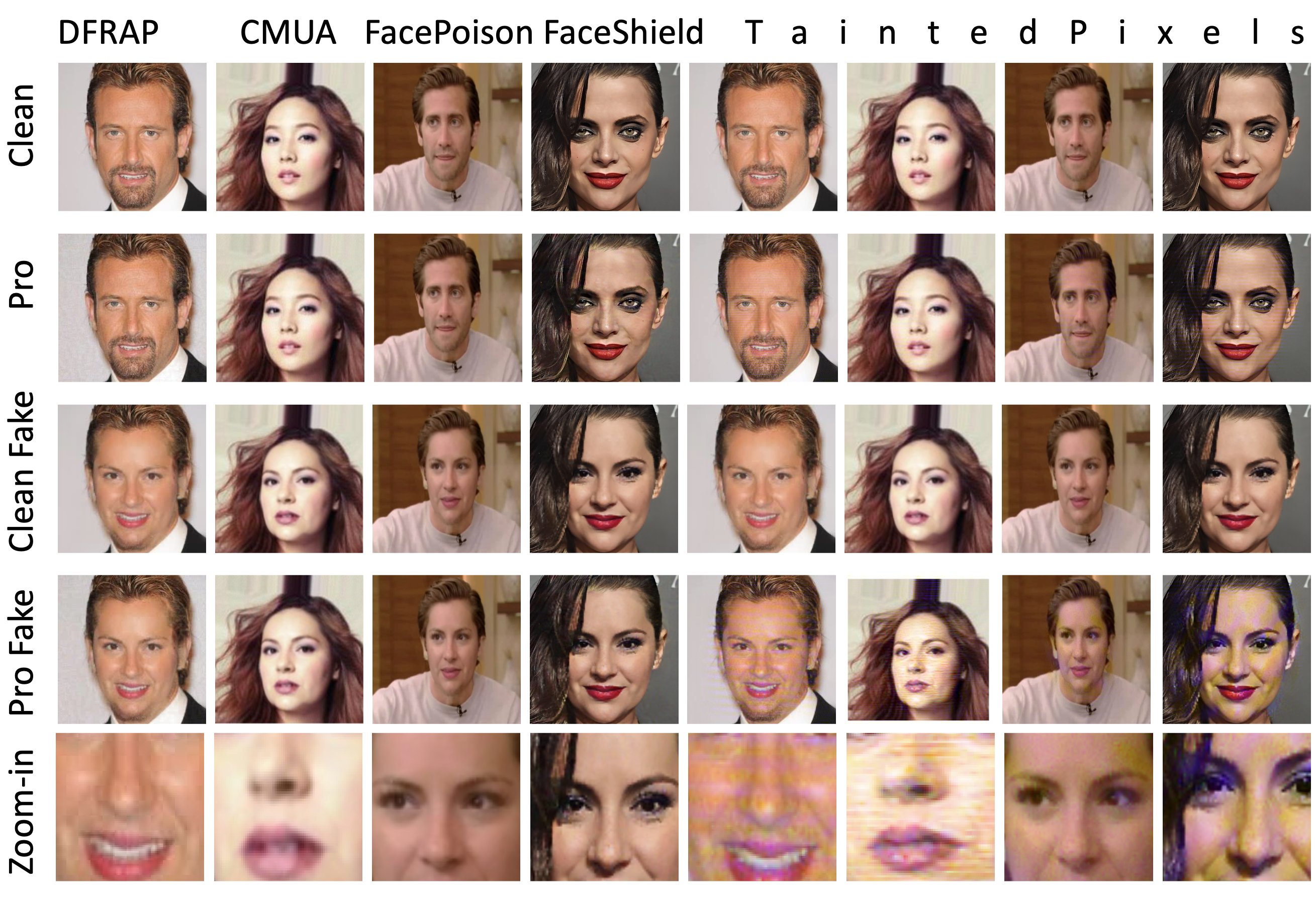}}
\caption{
Qualitative results. (R1 \& R2): the unprotected (clean) and protected source videos, resp. (R3): fakes generated from clean. (R4): fakes from protected. (R5): zoom in of (R4). \method{} creates obvious purple artifacts in every video frame, easily exposing the forgery to human eyes.
}
\label{fig:qualitative_results}
\end{minipage}\hfill


\end{figure*}

\textbf{Qualitative Results.} We use Roop as the manipulation tool and visualize some samples. Figure~\ref{fig:qualitative_results} shows that protected videos remain visually close to their clean sources. After manipulation, however, protected videos exhibit visible failures, typically manifested as yellowish and purplish color distortions in facial regions. Compared to rival protection methods, \method{} produces more noticeable artifacts in the videos forged from protected sources, thus making them easier to detect.

\begin{table*}[t]
\centering
\caption{
Comparison with SOTA methods (columns) under 3 manipulation tools (rows). 
D1 and D2 denote NPR and Deep-Fake-Detector-v2-Model, respectively. ``Clean'' denotes the unprotected setting, and the remaining columns correspond to forgeries generated from protected videos. The last three rows report the visual quality of the protected videos generated from each method.
}
\label{tab:quality_matched_main_comparison}
\resizebox{\textwidth}{!}{
\begin{tabular}{l l l c c c c c c}
\toprule
\hline
\multirow{2}{*}{Tool} & \multirow{2}{*}{Detector} & \multirow{2}{*}{Metric}
& \multirow{2}{*}{Clean}
& \multirow{2}{*}{DFRAP}
& \multirow{2}{*}{CMUA}
& \multirow{2}{*}{FacePoison}
& \multirow{2}{*}{FaceShield}
& \multirow{2}{*}{\method{} (ours)} \\
& & & & & & & &  \\
\midrule
\multirow{4}{*}{Roop}
& \multirow{2}{*}{D1 Fake}
& Fake Rate $\uparrow$ & 78.33 & 79.63 & 78.64 & 80.33 & 84.32 & \bf{89.12} \\
&
& Fake Conf. $\uparrow$ & 75.67 & 78.72  & 76.87 & 73.95& 80.32 & \bf{86.58} \\
\cmidrule(lr){2-9}
& \multirow{2}{*}{D2 Fake}
& Fake Rate $\uparrow$ & 82.43  & 84.51 & 83.02 & 83.93 & 86.71& \bf{90.27} \\
&
& Fake Conf. $\uparrow$ & 72.10  & 75.39 & 74.95 & 80.63 & 81.66 & \bf{86.92} \\
\midrule
\multirow{4}{*}{FaceFusion}
& \multirow{2}{*}{D1 Fake}
& Fake Rate $\uparrow$ & 80.87 & 80.39  & 82.01 & 82.20 & 84.02 & \bf{88.71} \\
&
& Fake Conf. $\uparrow$ & 73.43 & 76.74 & 75.43 & 74.93 & 79.34 & \bf{89.14} \\
\cmidrule(lr){2-9}
& \multirow{2}{*}{D2 Fake}
& Fake Rate $\uparrow$ & 80.29 & 82.31  & 83.34 & 83.72 & 85.48 & \bf{90.64}\\
&
& Fake Conf. $\uparrow$ & 72.93 & 78.46  & 75.54 & 77.23& 76.82& \bf{83.47} \\
\midrule
\multirow{4}{*}{MuseTalk}
& \multirow{2}{*}{D1 Fake}
& Fake Rate $\uparrow$ & 84.78 & 84.83  & 86.39 & 85.23 & 83.54 & \bf{93.68} \\
&
& Fake Conf. $\uparrow$ & 76.32 & 80.92  & 78.65 & 76.82 & 77.36 & \bf{86.46} \\
\cmidrule(lr){2-9}
& \multirow{2}{*}{D2 Fake}
& Fake Rate $\uparrow$ & 82.21 &  84.43 & 84.94 & 82.93 & 85.04 & \bf{92.92} \\
&
& Fake Conf. $\uparrow$ & 76.36 & 78.20  & 77.92 & 74.96 & 75.90 & \bf{90.21} \\
\midrule
\hline
\multirow{3}{*}{Protected videos}
& & PSNR $\uparrow$
& N/A & 30.23 & 37.43 & \bf{38.34} & 32.73 & 30.54 \\
& & SSIM $\uparrow$
& N/A & 0.8923 & 0.9342 & 0.9296 & \bf{0.9349} & 0.9103 \\
& & V-LPIPS $\downarrow$
& N/A & 0.1264 & 0.0527 & 0.0092 & 0.1941 & \bf{0.0042} \\
\bottomrule
\hline
\end{tabular}
}
\end{table*}

\subsection{Ablation Study}
\label{sec:ablation}

We ablate the main components of \method{} in
Table~\ref{tab:component_ablation}. The structured periodic prior tests whether
coherent blue-channel patterns are more effective than unstructured noise. The
high-frequency mask tests whether placing perturbations in textured regions
improves imperceptibility. Face-aware allocation tests whether using the
manipulation-sensitive facial region improves budget efficiency. The visibility
budget tests whether the method can preserve source-side naturalness while still
causing forged outputs to fail. Motion-adaptive scaling and video-level LPIPS
test the contribution of the video-aware deployment strategy. Results demonstrate that structured, region-aware, and temporally adaptive perturbations are all critical to effectiveness.


\begin{table*}[t]
\centering
\caption{
Component ablation of \method{}. We report average results over Roop,
FaceFusion, and MuseTalk. A better method should achieve high fake rate while
maintaining low source video-LPIPS.
}
\label{tab:component_ablation}
\resizebox{\textwidth}{!}{
\begin{tabular}{c c c c c c c c}
\toprule
Periodic Prior & HF Mask & Face-aware & Visibility Budget & Motion Scaling & Video LPIPS
& Avg. Fake Rate $\uparrow$ & Video-LPIPS $\downarrow$ \\
\midrule
-- & -- & -- & -- & -- & -- & 81.49 & -- \\
\checkmark & -- & -- & -- & -- & -- & 86.10 & 0.1094 \\
\checkmark & \checkmark & -- & -- & --& -- & 87.17  & 0.2036 \\
\checkmark & \checkmark & \checkmark & -- & -- & -- & 87.84 & 0.1895 \\
\checkmark & \checkmark & \checkmark & \checkmark & -- & -- & 88.35 & 0.0164 \\
\checkmark & \checkmark & \checkmark & \checkmark & \checkmark & -- & 89.62 & 0.0103 \\
\checkmark & \checkmark & \checkmark & \checkmark & \checkmark & \checkmark
& 90.89 & 0.0042 \\
\bottomrule
\end{tabular}
}
\end{table*}

\subsection{Human Perceptual Study}
\label{sec:human_study}

We further evaluate perceptual effects through a human study on Amazon Mechanical Turk. Forty non-expert participants viewed a randomly shuffled set of videos from four conditions (100 videos per condition): (1) \emph{clean source}, i.e., the videos sampled from FaceForensics++ and Celeb-DF; (2) \emph{protected source}, i.e., the output of \method{}; (3) \emph{forged from clean source}, i.e., a forgery generated by applying a public tool to the original unprotected video; and (4) \emph{forged from protected source}, i.e., a forgery generated by applying the same tool to our protected video. We selected 400 videos from our sample set as the stimuli, covering diverse lighting conditions, backgrounds, and skin tones, using Roop, FaceFusion, and MuseTalk as representative manipulation methods. Participants were asked whether each video appears suspicious or fake. For source videos, we report the fraction of affirmative responses as the \emph{false alarm rate}; for manipulated videos, we report it as the \emph{fake detection rate}. 



\begin{table}[t]
\centering
\small
\setlength{\tabcolsep}{5pt}
\caption{Human study results reported as the percentage of videos judged as fake. Lower values for source videos indicate better visual naturalness, while higher values for manipulated videos indicate that the forged outputs are easier for humans to recognize as fake.}
\label{tab:human_study}
\begin{tabular}{c c || l c c}
\toprule
Clean Source $\downarrow$ & Protected Source $\downarrow$ & Tool & Forged from Clean $\uparrow$ & Forged from Protected $\uparrow$ \\
\midrule
\multirow{4}{*}{0.80} & \multirow{4}{*}{3.26} & Roop & 54.23 & 89.42 \\
& & FaceFusion & 51.80 & 94.51 \\
& & MuseTalk & 64.11 & 88.24 \\
\cmidrule(lr){3-5}
& & Average & 56.71 & \textbf{90.72} \\
\bottomrule
\end{tabular}
\end{table}

As shown in Tab.~\ref{tab:human_study}, clean source videos receive 0.8\%, and protected source videos receive only 3.26\% false alarms, indicating that our perturbations do not noticeably affect visual quality. In contrast, forgeries from clean sources are identified as fake in 56.71\% of cases on average, reflecting the baseline detectability of current manipulation tools. When the same tools are applied to protected videos, the average detection rate increases to 90.72\%, a significant gain of 34.01\% points. This gain demonstrates that \method{} effectively helps humans detect forgery more readily.




\section{Conclusion}

To the best of 
our knowledge, \method{} is the first of its kind: a proactive defense for facial videos that operationalizes 
an asymmetric visibility trade-off; i.e.,~structured periodic perturbations 
in the blue channel of facial regions stay inconspicuous in the 
published video and become exposed only after manipulation. \method{} does not use any information about the manipulation tool, not even API access. To be sure, a 
sufficiently determined attacker aware of the perturbation could attempt to remove it through targeted denoising or 
heavy compression. We mitigate against this possibility indirectly: by 
keeping video-level LPIPS low, the published video is less likely to arouse suspicion even from attackers. We hope that our work encourages further research into proactive defenses.


\begin{ack}
Use unnumbered first level headings for the acknowledgments. All acknowledgments
go at the end of the paper before the list of references. Moreover, you are required to declare
funding (financial activities supporting the submitted work) and competing interests (related financial activities outside the submitted work).
More information about this disclosure can be found at: \url{https://neurips.cc/Conferences/2026/PaperInformation/FundingDisclosure}.

Do {\bf not} include this section in the anonymized submission, only in the final paper. You can use the \texttt{ack} environment provided in the style file to automatically hide this section in the anonymized submission.
\end{ack}

\bibliographystyle{plainnat}
\bibliography{references}


\appendix
\newpage
\section{Technical appendices and supplementary material}

\paragraph{Hyperparameter and auxiliary-term definitions.}
To improve reproducibility and avoid ambiguity in the notation, we summarize the
main initial hyperparameters used in our implementation in Table~\ref{tab:param_defs}.
We also list the auxiliary visibility-related quantities in
Table~\ref{tab:visibility_defs}. All experiments use the initial values in Table~\ref{tab:param_defs}. We fine-tune the parameters based on 6 frames.
\begin{table*}[!htbp]
\centering
\caption{
Summary of hyperparameters.
}
\label{tab:param_defs}
\resizebox{\textwidth}{!}{
\begin{tabular}{l l p{0.58\textwidth}}
\toprule
Symbol / Term & Value / Definition & Description \\
\midrule
$\alpha$ & $3.5$ & Amplitude of the structured blue-channel prior. \\
$r$ & $0.8$ & Fraction of high-frequency patches selected. \\
$p$ & $2.0$ & Period of the local periodic pattern. \\
$\epsilon$ & $0.25$ & Perturbation magnitude bound. \\
$\lambda_{\mathrm{d}}$ & $3.0$ & Weight for $\mathcal{L}_{\mathrm{d}}$. \\
$\lambda_{\mathrm{s}}$ & $1.0$ & Weight for $\mathcal{L}_{\mathrm{s}}$. \\
$\lambda_{\mathrm{v}}$ & $1.0$ & LPIPS perceptual budget. \\
$s_{\mathrm{in}}$ & $12$ & Perturbation scale inside the face region. \\
$s_{\mathrm{out}}$ & $0.8$ & Perturbation scale outside the face region. \\
$K$ & $120$ & Maximum number of Adam steps. \\
$\eta$ & $0.009$ & Adam learning rate. \\

\bottomrule
\end{tabular}
}
\end{table*}

\begin{table*}[!htbp]
\centering
\caption{
Definitions of auxiliary visibility-related quantities.
}
\label{tab:visibility_defs}
\resizebox{\textwidth}{!}{
\begin{tabular}{l p{0.72\textwidth}}
\toprule
Term & Role \\
\midrule
$C_b(x)$ &
Blue-difference chroma channel used to measure blue-channel color changes. \\

$\mathcal{B}_{\mathrm{vis}}(x)$ &
Content-adaptive stripe visibility budget estimated from clean-frame chroma statistics. \\
$W_{\mathrm{ws}}$ &
Combined soft white/skin-region mask for color-cast suppression. \\

$\mathcal{B}_{\mathrm{cast}}(x)$ &
Content-adaptive color-cast budget estimated from clean-frame white/skin chroma statistics. \\
$\mathrm{RowCoh}(\Delta C_b)$ &
Measures row-wise coherent chroma drift, corresponding to visible stripe/banding artifacts. \\

$\mathrm{LocalChr}(\Delta C_b)$ &
Measures local weighted blue-chroma deviation. \\
\bottomrule
\end{tabular}
}
\end{table*}


\end{document}